\documentclass[conference]{IEEEtran}
\IEEEoverridecommandlockouts
\usepackage{amsmath,amssymb,amsfonts}
\usepackage{algorithmic}

\usepackage{graphicx}
\usepackage{textcomp}
\usepackage{xcolor}
\usepackage{hyperref}
\hypersetup{
    colorlinks=true,
    allcolors=black
}
\usepackage[utf8]{inputenc}
\usepackage[T1]{fontenc}
\usepackage{siunitx}
\def\BibTeX{{\rm B\kern-.05em{\sc i\kern-.025em b}\kern-.08em
    T\kern-.1667em\lower.7ex\hbox{E}\kern-.125emX}}
\usepackage{cleveref}
\usepackage{lipsum}

\usepackage{pgfplots}
\pgfplotsset{compat=1.18}
\usepackage{pgf}
\usepackage{tikz}
\usetikzlibrary{arrows.meta, positioning}

\usepackage{mathtools}

\usepackage{pgfplots}
\usetikzlibrary{arrows.meta}
\usetikzlibrary{shapes.geometric}
\usetikzlibrary{positioning}
\usepackage[outline]{contour}
\pgfplotsset{compat=1.18}
\usepackage{overpic}
\usepackage{makecell}
\usepackage{multirow}
\usepackage{booktabs}

\usepackage[style=ieee, backend=biber, giveninits=true, maxnames=5]{biblatex}
\AtEveryBibitem{%
    \clearfield{month}%
    \clearfield{language}
    \clearfield{urlyear}%
    \clearfield{issn}%
    \clearfield{doi}
    \clearfield{url}
}

\defbibenvironment{bibliography}
  {\list
     {\printfield[labelnumberwidth]{labelnumber}}
     {%
       \setlength{\labelwidth}{2em}
       \setlength{\labelsep}{0.7em}  
       \setlength{\leftmargin}{\dimexpr\labelwidth+\labelsep\relax}
       \setlength{\itemsep}{0.1\baselineskip}
       \setlength{\parsep}{0pt}
     }}
  {\endlist}
  {\item}

\DeclareBibliographyDriver{inproceedings}{%
  \printnames{author}%
  \setunit{\labelnamepunct}\newblock
  \printfield[title]{title}%
  \newunit
  \printfield{booktitle}%
  \iffieldundef{volume}{}{\setunit{\addcomma\space}\printfield{volume}}%
  \iffieldundef{pages}{}{\setunit{\addcomma\space}\printfield{pages}}%
  \setunit{\addcomma\space}\printfield{year}%
  \newunit\newblock
}

\DeclareBibliographyDriver{techreport}{%
  \printnames{author}%
  \setunit{\labelnamepunct}\newblock
  \printfield[title]{title}%
  \newunit
  \printfield{institution}%
  \iffieldundef{number}{}{\setunit{\addcomma\space}\printfield{number}}%
  \iffieldundef{pages}{}{\setunit{\addcomma\space}\printfield{pages}}%
  \setunit{\addcomma\space}\printfield{year}%
  \newunit\newblock
}

\def\BibTeX{{\rm B\kern-.05em{\sc i\kern-.025em b}\kern-.08em
    T\kern-.1667em\lower.7ex\hbox{E}\kern-.125emX}}
    
\begin{document}

\title{Temperature-Aware Heat Pump Modeling for Large-Scale Energy System Optimization
\thanks{This work is part of the project RINGs (grant number FO999905702), which has received funding within the framework of Energieforschung, a research and technology programme of the Klima- und Energiefonds (Austria).}
}

\author{
\IEEEauthorblockN{Simon Malacek$^{a,b}$, Sonja Wogrin$^{a,b}$, Yannick Werner$^{a,b}$}
\IEEEauthorblockA{
$^a$\textit{Institute of Electricity Economics and Energy Innovation, Graz University of Technology}, Graz, Austria\\
$^b$\textit{Research Center ENERGETIC, Graz University of Technology}, Graz, Austria}
\IEEEauthorblockA{\href{mailto:simon.malacek@tugraz.at}{simon.malacek@tugraz.at}, 
\href{mailto:wogrin@tugraz.at}{wogrin@tugraz.at},
\href{mailto:yannick.werner@tugraz.at}{yannick.werner@tugraz.at}}
}

\maketitle

\newcommand\copyrighttext{%
  \footnotesize \textcopyright \the\year{} IEEE. Personal use of this material is permitted. Permission from IEEE must be obtained for all other uses, including reprinting/republishing this material for advertising or promotional purposes, collecting new collected works for resale or redistribution to servers or lists, or reuse of any copyrighted component of this work in other works.}

\newcommand\copyrightnotice{%
\begin{tikzpicture}[remember picture,overlay]
\node[anchor=south,yshift=10pt] at (current page.south) {\fbox{\parbox{\dimexpr0.75\textwidth-\fboxsep-\fboxrule\relax}{\copyrighttext}}};
\end{tikzpicture}%
}

\newcommand\IEEEcopyrighttext{\textbf{979-8-3195-3554-2/26/\$31.00 ©2026 IEEE}}
\newcommand\IEEEcopyrightnotice{%
\begin{tikzpicture}[remember picture,overlay]
\node[anchor=south west, xshift=1.6cm, yshift=1.3cm] 
at (current page.south west) {%
 \IEEEcopyrighttext};
\end{tikzpicture}%
}

\copyrightnotice 
\vspace{-10pt}

\begin{abstract}
Heat pumps are expected to dominate the heating sector, substantially increasing peak electricity demand. At the same time, building thermal inertia enables operational strategies, providing temporal flexibility in heat pump operation and short-term demand response. However, this dynamic behavior is not yet represented in large-scale energy system optimization models. To address this gap, we present an innovative formulation of building thermal inertia. The resulting temperature variable is integrated into a novel conic temperature-aware heat pump efficiency formulation, enabling a more precise emulation of smart control strategies. In a case study of the European energy system, we show that the approach captures operational heating flexibility while remaining computationally efficient. The results indicate substantial untapped flexibility potential, enabling up to a~22\% reduction in heating-related electricity costs. This potential can be realized through a suitable energy market design that incentivizes coordinated heat pump control, individually or via aggregators.
\end{abstract}


\section{Introduction}
Large-scale energy system optimization models (ESOMs), such as  PyPSA~\cite{horsch_pypsa-eur_2018} or LEGO~\cite{wogrin_lego_2022}, are essential tools for supporting decision-making in energy system planning and market design. In practice, modelers face a trade-off between representing technical details and ensuring computational tractability. While for modeling the electricity system, different formulations with levels of technical complexity (e.g.\ transport models, DC or AC power flow, or unit commitment with ramping constraints)~\cite{kotzur_modelers_2021} are commonly available, such a granularity of formulations is largely lacking for the heat sector. 

A key gap in heat sector modeling is the accurate representation of technical and operational details, particularly heat transfer physics, thermal inertia (TI), and the temperature-dependent performance and control strategies of heat pumps (HPs). It is expected that, by 2050, around 30\% of the total heating demand will be covered via individual HPs in rural areas, with an additional 20\% supplied by HPs via district heating networks within the European Union~\cite{noauthor_hre5_nodate}. In addition, their electricity consumption directly couples the heat and power sectors, unlike other heating technologies such as biomass. This coupling may challenge power system operation due to additional demand peaks, but, at the same time, opens up substantial demand response potential, as potentially partially dispatchable electricity demand~\cite{lyons_clean_2023, muller_large-scale_2019, sperber_reduced-order_2020,freischlad_harnessing_2026}.
Providing demand response requires flexibility in HP operation, which can be achieved by exploiting the TI of buildings. This allows short-term shifts in heat supply with only small deviations in indoor temperature and, consequently, thermal comfort.
To realize these demand response potentials in electricity markets, a suitable market design is required that incentivizes individuals to operate HPs optimally with respect to the system, for example, through variable electricity prices~\cite{schachter_business_2015} or through aggregators~\cite{semmelmann_aggregator_2026} that take over HP operation and provide thermal comfort as a service.

To understand the system-wide implications of power-system-optimal HP operation in the first place, it is crucial that energy system models accurately represent (i) building TI effects and (ii) HP conversion efficiency. We briefly introduce both aspects in the following.

\subsection{Thermal building inertia modeling}
To model the TI of buildings, a virtual heat storage (VHS)~\cite{noauthor_heat_nodate} is often introduced that mimics the heat energy stored as a temperature deviation from the building setpoint level~\cite{askeland_low-parameter_2023}. Temperature deviations beyond a predefined comfort range are then penalized, as for example in~\cite{semmelmann_aggregator_2026}. Similarly, \cite{lu_thermal_2020} derive a parameterization from an aggregated building TI model, representing the entire building as a battery-like thermal storage. In \cite{wang_integrated_2023}, TI is likewise represented as a storage for the whole building, combined with particle swarm optimization beyond standard linear optimization approaches. 
A key drawback of these simplified VHS representations is their limited ability to capture the physical characteristics of heat transfer processes. Moreover, they rely heavily on the definition of temperature deviation penalties and predefined comfort windows, which can be difficult to parameterize.


\subsection{Temperature-aware heat pump conversion efficiency}
In most optimization models, the coupling between the power and heat sectors via HPs is represented by a parameter, typically either a constant~\cite{wiese_balmorel_2018},~\cite{franken_power_2025} or an hourly pre-calculated, outdoor-temperature-dependent~\cite{ruhnau_time_2019},~\cite{ruhnau_empirical_2023} coefficient of performance (COP). This formulation links HP electricity consumption directly to the heat demand time series. However, the model is not aware of the supply temperature in the heat distribution system, i.e., the HP sink temperature, affecting HP efficiency~\cite{huchtemann_simulation_2013},~\cite{maier_assessing_2022}. 
The motivation for sink-temperature-aware HP efficiency stems from several considerations in heat engineering. Maier et~al.~\cite{maier_assessing_2022} show that thermal preheating or storage charging requires higher supply temperatures, which in turn reduces HP efficiency. Conversely, \cite{huchtemann_simulation_2013} demonstrate that optimizing the supply temperature can yield energy savings of around 7\%. However, both the model predictive control approach and the detailed simulation rely on nonlinear and nonconvex relationships, restricting scalability to large-scale ESOMs.

\subsection{Original contribution}
Both aspects outlined above have been recognized in the literature and are partly addressed in small-scale models focusing on individual buildings or energy hubs (e.g.~\cite{askeland_low-parameter_2023}). However, these formulations are typically not embedded in large-scale ESOMs. This may lead to a misestimation of HP flexibility -- either overestimating it by neglecting efficiency losses at higher supply temperatures or underestimating it by insufficiently capturing building TI.

In this work, we address these methodological gaps by proposing a formulation for large-scale ESOMs that (i) more accurately represents the thermal storage potential of buildings by separating TI contributions between the floor, acting as the heat delivery system, and the room itself, thereby overcoming limitations of simple VHS models; and (ii) employs a conic, convex temperature-aware HP conversion formulation derived from a detailed physical model that captures supply-temperature-dependent effects; and (iii) couples this temperature awareness to the floor temperature, representing part of the VHS and the short-term heating history of the building. Together, this enables a more physically accurate estimation of actual \emph{thermo-temporal} flexibility potentials compared to existing formulations. We demonstrate this through a stylized case study of the European energy system, analyzing flexibility provision, peak electricity demand of HPs, and thermal discomfort. Additionally, we show the computational efficiency and scalability of the proposed formulations for large-scale ESOMs.

The remainder of the paper is organized as follows: Section~\ref{sc:model_formulation} introduces the proposed formulations. Section~\ref{sc:case_study} presents the case study, followed by a discussion of the key results in Section~\ref{sc:results}. Finally, Section~\ref{sc:conclusion} outlines the implications, applicability, and directions for future research.

\section{Model formulation}
\label{sc:model_formulation}
After briefly outlining the integration into a large-scale ESOM in Section~\ref{sc:sector_coupling}, Section~\ref{sc:therm_building_model} presents approaches to modeling building TI, followed by Section~\ref{sc:heat_pump_model}, which introduces a parametric and a temperature-aware HP formulation.

\subsection{Sector coupling}
\label{sc:sector_coupling}
The formulations introduced here are intended for large-scale ESOMs and are compatible with common representations of the power sector. The following formulation illustrates the sector coupling from power to heat via HPs. To this end, the heating sector within a given area is represented by a set of heat nodes~$n \in \mathcal{N}$, which are uniquely assigned to electrical nodes~$i \in \mathcal{I}$ of the power system. For each electrical node~$i$, the subset~$\mathcal{N}_i \subseteq \mathcal{N}$ denotes all heat nodes connected to node~$i$. Each heat node~$n$ is characterized by an exogenous heat demand~$Q^{\mathrm{d}}_{n,h}$ for each time step~$h \in \mathcal{H}$. For energy conversion, the ensemble of HPs installed at heat node~$n$ generates heat~$q^{\mathrm{hp}}_{n,h}$ in each time step as a function of its electrical power consumption~$p^{\mathrm{hp}}_{n,h}$.

The electricity demand induced by HPs, which has to be supplied by the power system, is given by:
\begin{equation}
\hat{p}^{\mathrm{hp}}_{i,h} = \sum_{n \in \mathcal{N}_i} p^{\mathrm{hp}}_{n,h}
\quad \forall i \in \mathcal{I},~\forall h \in \mathcal{H}.
\end{equation}

\subsection{Thermal building inertia model}
\label{sc:therm_building_model}
In the following, we introduce the concepts for modeling the thermal storage function of buildings arising from inherent TI. We focus on HPs without backup heating and underfloor heating as the dominant distribution system.

\paragraph{No storage model}
If no TI of buildings is considered, heat generation must exactly match heat demand in every time step:
\begin{equation}
Q^{\mathrm{d}}_{n,h} = q^{\mathrm{hp}}_{n,h}
\quad \forall n \in \mathcal{N},~\forall h \in \mathcal{H}.
\end{equation}
This formulation provides no temporal flexibility in electricity consumption for heating and therefore constitutes an inflexible temperature control strategy at the building level.

\paragraph{Simple storage model}
To account for building TI in a simplified manner, a VHS using a conventional storage formulation can be modeled, as, e.g., done in~\cite{lu_thermal_2020, semmelmann_aggregator_2026}. For this purpose, we introduce nonnegative variables for the storage level~$q^{\mathrm{lv}}_{n,h}$, storage charging~$q^{\mathrm{ch}}_{n,h}$, and storage discharging~$q^{\mathrm{dch}}_{n,h}$. 
Formulating a simple storage model the heat energy stored in time step $h$ is captured by:
\begin{equation}
q^{\mathrm{lv}}_{n,h+1} = q^{\mathrm{lv}}_{n,h} - q^{\mathrm{dch}}_{n,h} + q^{\mathrm{ch}}_{n,h}
\quad \forall n \in \mathcal{N},~\forall h \in \mathcal{H}.
\end{equation}
Where the heat balance is then written as:
\begin{equation}
Q^{\mathrm{d}}_{n,h} = q^{\mathrm{hp}}_{n,h} + q^{\mathrm{dch}}_{n,h} - q^{\mathrm{ch}}_{n,h}
\quad \forall n \in \mathcal{N},~\forall h \in \mathcal{H}.
\end{equation}

The storage level can be directly related to the average building ensemble temperature~$t^{\mathrm{b}}_{n,h}$ (all temperatures in Kelvin) by
\begin{equation}
q^{\mathrm{lv}}_{n,h} = C^{\mathrm{b}}_n \cdot t^{\mathrm{b}}_{n,h}
\quad \forall n \in \mathcal{N},~\forall h \in \mathcal{H},
\label{eq:q2t_relation}
\end{equation}
where the heat capacity~$C^{\mathrm{b}}_n$ denotes the aggregated total heat capacity of the buildings at heat node~$n$.


Deviations of the indoor building temperature from the setpoint $T^{\mathrm{b,set}}_n$ lead to thermal discomfort. To account for this in the optimization, a discomfort cost term $k^\mathrm{discom}$ is introduced and added to the overall objective function.

The penalty scales with the exposed area $A_n$ and a cost coefficient per unit temperature deviation and nominal area, $K^{\mathrm{penalty}}$. Using the relation between storage level and indoor temperature given in~\eqref{eq:q2t_relation}, the hourly building temperature is expressed via the thermal storage level. The resulting formulation reads:

\begin{equation}
k^\mathrm{discom} = \sum_{h \in \mathcal{H}, n\in\mathcal{N}} {A_n}
\left| \frac{q^{\mathrm{st,lv}}_{n,h}}{C^\mathrm{b}} - T^{\mathrm{b,set}}_n \right| \cdot K^{\mathrm{penalty}}.
\end{equation}

This approach can be extended to more sophisticated penalization concepts (e.g., \cite{semmelmann_aggregator_2026}, \cite{ashouri_optimal_2013}) without impairing the formulations introduced here.

While this formulation allows temporal flexibility in heat supply, any use immediately causes room temperature deviations and penalty costs, making the model highly sensitive to parameter choices, difficult to calibrate, and necessitating the assumption of no occupant control over indoor temperatures.

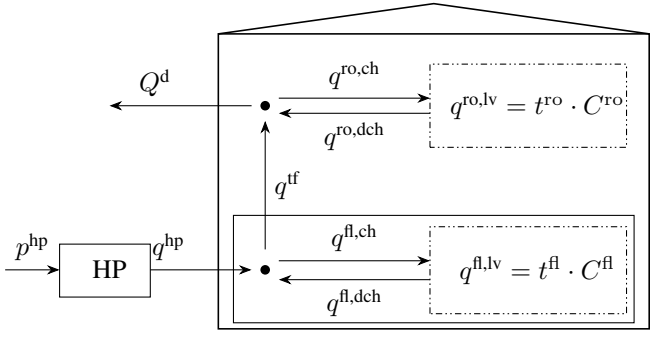
\begin{figure}
    \centering

\begin{tikzpicture}[
    >=Stealth,
]
\draw[semithick] (2.1, 0) rectangle (7.8, 3.9);
\draw[semithick] (2.1, 3.9) -- (4.95, 4.3) -- (7.8, 3.9);

\draw[thin] (2.3, 0.08) rectangle (7.6, 1.5);

\draw[thin, dash pattern=on 2.4pt off 1.2pt on 0.6pt off 1.2pt on 0.6pt off 1.2pt] (4.9, 2.4) rectangle (7.50, 3.5);
\node at (6.32, 2.95) {$q^{\text{ro,lv}} = t^\mathrm{ro} \cdot  C^\mathrm{ro}$};

\draw[thin, dash pattern=on 2.4pt off 1.2pt on 0.6pt off 1.2pt on 0.6pt off 1.2pt] (4.9, 0.2) rectangle (7.50, 1.35);
\node at (6.32, 0.78) {$q^{\text{fl,lv}} = t^\mathrm{fl} \cdot  C^\mathrm{fl}$};

\fill (2.72, 2.95) circle (1.8pt);
\fill (2.72, 0.78) circle (1.8pt);

\draw[->, thin] (2.9, 3.05) -- node[above] {$q^{\text{ro,ch}}$} (4.9, 3.05);
\draw[<-, thin] (2.9, 2.85) -- node[below] {$q^{\text{ro,dch}}$} (4.9, 2.85);

\draw[->, thin] (2.9, 0.9) -- node[above] {$q^{\text{fl,ch}}$} (4.9, 0.9);
\draw[<-, thin] (2.9, 0.65) -- node[below] {$q^{\text{fl,dch}}$} (4.9, 0.65);

\draw[<-, thin] (0.66, 2.95) -- node[above, pos=0.32] {$Q^{\text{d}}$} (2.54, 2.95);

\draw[->, thin] (2.72, 1.05) -- node[right] {$q^{\text{tf}}$} (2.72, 2.75);

\draw[thin] (0, 0.42) rectangle (1.20, 1.12);
\node at (0.66, 0.78) {HP};

\draw[->, thin] (-0.72, 0.78) -- node[above] {$p^{\text{hp}}$} (0, 0.78);

\draw[->, thin] (1.20, 0.78) -- node[above, pos=0.18] {$q^{\text{hp}}$} (2.54, 0.78);

\end{tikzpicture}
    \vspace{-0.85cm}
    \caption{Advanced storage model illustrating energy flows in a virtual aggregated building.}
    \label{fig:advanced_storage_model}
\end{figure}

\paragraph{Advanced storage model}
The advanced storage model proposed here is motivated by the increasing use of underfloor heating in residential buildings. In such systems, a large share of the effective TI is associated with the floor structure -- typically concrete with high mass and heat capacity -- rather than with the room air and remaining building components, which have comparatively low heat capacity~\cite{verbeke_thermal_2018, li_improving_2020}. We therefore distinguish between the heat capacity of the actively heated building mass, $C^{\mathrm{fl}}$, and that of the room, $C^{\mathrm{ro}}$.\footnote{For comparability within the case study, we impose $C^{\mathrm{b}}_n = C^{\mathrm{fl}}_n + C^{\mathrm{ro}}$. } The total heat storage is then modeled as two separate but coupled components, as illustrated in Fig.~\ref{fig:advanced_storage_model}. We do this by introducing a VHS level for the floor,~$q^\mathrm{fl,lv}$, and the room,~$q^\mathrm{ro,lv}$, and again assume that they are proportional to the respective temperatures,~$t^\mathrm{fl}$ and~$t^\mathrm{ro}$, such that~$q^\mathrm{fl,lv} = t^\mathrm{fl} \cdot C^\mathrm{fl}$ and~$q^\mathrm{fl,lv} = t^\mathrm{ro} \cdot C^\mathrm{ro}$, analogously to~\eqref{eq:q2t_relation}. 
We explain the room, transfer, and floor components in detail below.

The \textbf{room} heat storage level is represented similarly to the one in the simple building storage model: 
\begin{equation}
q^{\mathrm{ro,lv}}_{n,h+1} = q^{\mathrm{ro,lv}}_{n,h} - q^{\mathrm{ro,dch}}_{n,h} + q^{\mathrm{ro,ch}}_{n,h}
\quad \forall n \in \mathcal{N},~\forall h \in \mathcal{H},
\end{equation}
whereas the room heat balance reads:
\begin{equation}
D^{\mathrm{heat}}_{n,h} = q^{\mathrm{tf}}_{n,h} + q^{\mathrm{ro,dch}}_{n,h} - q^{\mathrm{ro,ch}}_{n,h}
\quad \forall n \in \mathcal{N},~\forall h \in \mathcal{H}.
\end{equation}

In contrast to directly equating heat generation from the HP with heat demand, heat is now first explicitly transferred from the underfloor heating system to the room via the auxiliary variable~$q^{\mathrm{tf}}_{n,h}$. Following the approach from~\cite{soller_heizungsbauer_1998}, this heat \textbf{transfer} is driven by the temperature gradient between floor and room, expressed here through the respective storage levels, which are proportional to temperatures according to~\eqref{eq:q2t_relation}, and by the product of a heat transfer coefficient~$\alpha$ and the total floor area~$A_n$ at heat node~$n$:
\begin{equation}
q^{\mathrm{tf}}_{n,h} = A_n \cdot \alpha \cdot 
\left(
\frac{q^{\mathrm{fl,lv}}_{n,h}}{C^{\mathrm{fl}}_n}
-
\frac{q^{\mathrm{ro,lv}}_{n,h}}{C^{\mathrm{ro}}_n}
\right).
\end{equation}

Analogously to the room, the \textbf{floor} heat storage level is given by:
\begin{equation}
q^{\mathrm{fl,lv}}_{n,h+1} = q^{\mathrm{fl,lv}}_{n,h} - q^{\mathrm{fl,dch}}_{n,h} + q^{\mathrm{fl,ch}}_{n,h}
\quad \forall n \in \mathcal{N},~\forall h \in \mathcal{H},
\end{equation}
whereas the floor heat balance reads:
\begin{equation}
q^{\mathrm{hp}}_{n,h} = q^{\mathrm{tf}}_{n,h} - q^{\mathrm{fl,dch}}_{n,h} + q^{\mathrm{fl,ch}}_{n,h}
\quad \forall n \in \mathcal{N},~\forall h \in \mathcal{H}.
\end{equation}
Similarly to  the simple storage formulation, deviation from the room temperature setpoint $T^{\mathrm{ro,set}}_{n}$ is penalized in the objective function:
\begin{equation}k^\mathrm{discom} = \sum_{h \in \mathcal{H}, n \in \mathcal{N}} {A_n}
\left| \frac{q^{\mathrm{ro,lv}}_{n,h}}{C^\mathrm{ro}} - T^{\mathrm{ro,set}}_{n} \right| \cdot K^{\mathrm{penalty}}.
\end{equation}

In this formulation, however, no penalty is imposed on the floor storage level, as deviations do not affect thermal comfort. Instead, the floor storage level is bounded by operational constraints:
\begin{equation}
q^{\mathrm{fl,lv}}_{n,h} \le T^{\mathrm{fl,max}}_n \cdot C^{\mathrm{fl}}_n \quad\forall n \in \mathcal{N}, \forall h \in \mathcal{H} ,
\end{equation}
where~$T^{\mathrm{fl,max}}_n$ is determined by building standards.

This formulation is governed by actual physics, ensuring physically feasible operation of the VHS. By explicitly controlling the floor temperature, the floor is utilized as the primary thermal energy storage, resembling a more advanced building-level control strategy. Crucially, this is the only formulation that introduces the floor temperature as the HP sink, thereby enabling a fully coupled temperature-aware HP modeling approach.

\subsection{Heat pump model}
\label{sc:heat_pump_model}
In the following, we introduce two distinct HP formulations, starting with the state-of-the-art representation as a reference and subsequently presenting our proposed temperature-aware formulation. As we consider large ensembles of small HPs, we abstract from unit-level effects such as start-up behavior. For all cases, we consider a heat production limit by the installed capacity~$Q^{\mathrm{inst}}_n$.

\paragraph{Temperature-naive formulation}
In a simplified approach, an ensemble of HPs at heat node~$n$ can be described by:
\begin{equation}
q^{\mathrm{hp}}_{n,h} = \mathrm{COP}_{n,h} \cdot p^{\mathrm{hp}}_{n,h}
\quad \forall n \in \mathcal{N},~ \forall h \in \mathcal{H},
\end{equation}
where the $\mathrm{COP}_{n,h}$ is precomputed based on, for example, heat source and sink temperatures, as in~\cite{ruhnau_time_2019}, and is therefore treated as a fixed parameter. This approach assumes predefined temperature levels in the heat supply system, limiting the representation of advanced supply-temperature-aware control strategies that could increase HP efficiency.

\paragraph{Temperature-aware formulation}
To mimic advanced HP control strategies, a formulation that accounts for the current sink temperature is essential. Building on the advanced TI model, the floor temperature can be identified as the sink temperature, thereby additionally capturing temperature dynamics over time. 

To this end, we introduce a model motivated by the physics of the heating system, which describes HP efficiency as a function of the temperature levels, as well as the heat and mass transfer from the HP to the floor. The detailed derivation, parameter values, and assumptions are provided in Appendix~\ref{app:model_pararmetrization}. This approach allows us to derive a normalized performance map of heat output for each outdoor temperature. To represent this performance data in the optimization model, we fit Equation~\eqref{eq:fitting_cunction} with the fitting parameters $F$, $B$, and $M$, where the parameter $F$ is scaled by the installed thermal capacity $Q^\mathrm{inst}$ to allow generalization to arbitrary power ranges. Indices $n$ and $h$ are omitted for clarity.
\begin{equation}
\label{eq:fitting_cunction}
q^{\mathrm{hp}} = \frac{F}{Q^{\mathrm{inst}}} \cdot 
\frac{(p^{\mathrm{hp}})^2}{M - t^{\mathrm{fl}}} 
+ B \cdot p^{\mathrm{hp}}.
\end{equation}

\begin{figure}
    \centering
    \includegraphics{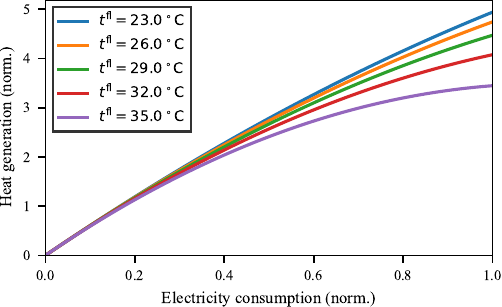}
    \caption{Normalized performance map of HP heat generation as a function of electricity consumption and floor temperature $t^\mathrm{fl}$.}
    \label{fig:normalised_performance_map}
\end{figure}

This novel formulation avoids the explicit modeling of intermediate temperature levels and mass flow rates while retaining the dominant physical effects. The fitted performance map, shown in Fig.~\ref{fig:normalised_performance_map}, achieves a coefficient of determination of~$R^2 = 0.985$.

For implementation in the large-scale ESOM, the floor temperature is replaced by $t^{\mathrm{fl}} = {q^{\mathrm{fl,lv}}} / C^{\mathrm{fl}}$ and the auxiliary variable:
\begin{equation}
s = \frac{(p^{\mathrm{hp}})^2}{M - q^{\mathrm{fl,lv}}/C^{\mathrm{fl}}}
\end{equation}
is introduced. 

This yields the following set of model constraints for each $n \in \mathcal{N}$ and $h \in \mathcal{H}$:
\begin{subequations}
\begin{align}
q^{\mathrm{hp}}_{n,h} = \frac{F_{n,h}}{Q^{\mathrm{inst}}} \cdot s_{n,h} 
+ B_{n,h} \cdot p^{\mathrm{hp}}_{n,h},
\label{conic_a} \\[0.5em]
\left(
M_{n,h} - q^{\mathrm{fl,lv}}_{n,h}/C^{\mathrm{fl}}_n 
\right) \cdot s_{n,h} 
\geq (p^{\mathrm{hp}}_{n,h})^2.
\label{conic_b}
\end{align}
\end{subequations}

Constraint~\eqref{conic_a} is linear, while~\eqref{conic_b} represents a rotated second-order cone constraint~\cite{aps_mosek_nodate} for~$A<0$ and~$M - {q^{\mathrm{st,floor,lv}}_{n,h}}/{C^{\mathrm{floor}}_n} > 0$. Such constraints can be solved efficiently or reformulated as a linear relaxation~\cite{raheli_conic_2023}, rendering the proposed formulation suitable for large-scale ESOMs.

Qualitatively, this formulation captures HP efficiency reductions associated with higher supply temperatures from two perspectives: (i) elevated floor temperatures, for example, due to prior heating, which are only represented through the temporal coupling introduced by the advanced TI model; and (ii) higher operating power levels, which require increased supply temperatures to drive heat transfer. Accounting for both effects yields a more physically consistent representation of HP operation in ESOMs.

\subsection{Summary}
We combine the introduced TI and HP formulations into the four heating-sector variants summarized in Table~\ref{tab:scenarios}. These formulations are general and can be integrated into any ESOM.

\begin{table}
    \centering
    \caption{Overview of the model variants.}
    \label{tab:scenarios}
    \begin{tabular}{c|c|c}
        Model & TI model & HP model  \\ \hline
        M1 & no  & temp-naive \\
        M2 & simple & temp-naive \\
        M3 & advanced & temp-naive \\
        M4 & advanced & temp-aware
    \end{tabular}
\end{table}

\section{Case study}
\label{sc:case_study}
We validate the four model variants from Table~\ref{tab:scenarios} by integrating them into the linear ESOM LEGO~\cite{wogrin_lego_2022}. There, the power sector is represented as an operational problem with a single node per bidding zone and limited transmission capacity between zones. It comprises 63 power zones, to which installed generation capacities are aggregated and assigned accordingly. The optimization is performed with hourly resolution over 8{,}760 time steps.
Data are taken from the TYNDP~\cite{entog_download_2023}, using the weather year~2008 and installed generation capacities for 2040. Time series for heat demand and COP are derived from the When2Heat dataset~\cite{ruhnau_time_2019} for the same weather year. A HP share of~30\% by 2040 is assumed~\cite{noauthor_hre5_nodate}.
To ensure a fair comparison, all models were solved using Gurobi~13.0.0 with the barrier method and without crossover on a workstation equipped with an Intel i9-12900 CPU (2.4~GHz, 16 cores) and 128~GB RAM. After presolve, the model comprised approximately 24\,M rows, 23\,M columns, and 74\,M nonzeros.

\section{Results}
\label{sc:results}
For comparison, key performance indicators (KPIs) are reported in Table~\ref{tab:scenario_comparison}. As expected, increasing system flexibility reduces total system costs and the average electricity cost for HPs in M2 and M3 by around 12\% compared to the inflexible M1. M4 further reduces total system costs and the average electricity cost for HPs by 22\% through improved supply temperatures, as the non-linear COP formulation enhances HP efficiency at low operating power levels (see Fig.~\ref{fig:normalised_performance_map}). 

While M1, by design, shows no temperature deviations, M2 and M3 exhibit similar values across all metrics. In contrast, M4 reduces deviations by favoring part-load operation and lower supply temperatures rather than fully exploiting TI. 
Annual average COP values reflect this behavior: M4 improves HP performance to $4.06$, compared to approximately $3.6$ in the other models, which operate more rigidly. Peak electrical demand of HPs increases across all flexible scenarios by about 30\% (M2 and M4) to 54\% (M3) relative to M1, likely due to the exploitation of lower electricity prices.

Finally, adding the simple storage formulation in M2 and the advanced storage formulation in M3 increases solver work\footnote{Gurobi metric for deterministic computational effort spent for solving.} by 28\% and 77\%, respectively. In contrast, the conic constraint formulation for the HP in M4 is computationally more efficient, reducing solver work by 11\% compared to the linear HP formulation in M3.

\begin{table}
\centering
\caption{KPIs for the four analyzed scenarios. All metrics were calculated across all heat nodes.}
\label{tab:scenario_comparison}
\begin{tabular}{p{3.87cm}cccc}
\hline
\textbf{Metric} & \textbf{M1} & \textbf{M2} & \textbf{M3} & \textbf{M4} \\
\hline
Total system costs in B€  & 49.5 & 48.9 & 48.9 & 48.5 \\
Avg. el. cost for HPs in €/MWh & 5.75 & 5.07 & 5.05 & 4.51 \\
Rel. change compared to M1 in \% & 0.0 & $-11.8$ & $-12.2$ & $-21.5$ \\
Max. positive temp. dev. in °C & 0.00 & 2.15 & 1.76 & 1.46 \\
Max. negative temp. dev. in °C & 0.00 & 0.82 & 0.82 & 0.70 \\
Annual COP  & 3.58 & 3.62 & 3.61 & 4.06 \\
Peak el. demand of HPs in GW & 95.8 & 126.9 & 147.2 & 124.6 \\
Solver work in arb. units & 1227 & 1634 & 2262 & 2007 \\
\hline
\end{tabular}
\end{table}

\begin{figure}
    \centering
    \includegraphics{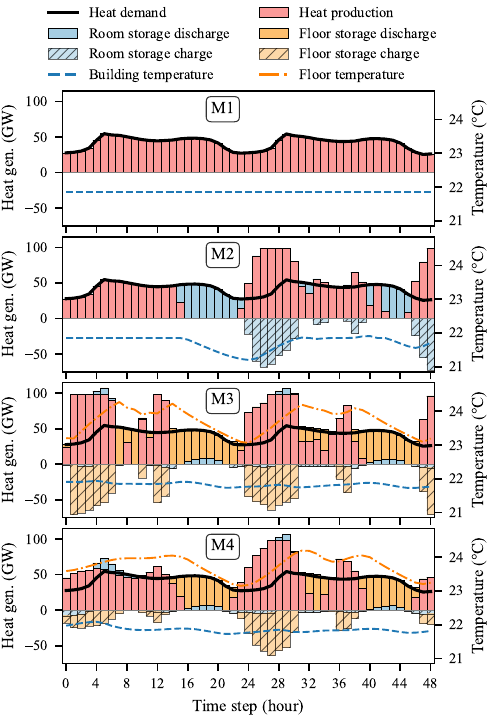}
    \caption{Comparison of the four scenarios. Heat demand and heat supply for the German heat node over a 48-hour period, starting at midnight on 12~December~2008.}
    \label{fig:scenario_comparison}
\end{figure}

For a qualitative comparison, two consecutive days of the German heat node are shown in Fig.~\ref{fig:scenario_comparison}. In M1, heat production strictly follows demand, with no temporal shifting. M2, by contrast, a negative temperature deviation occurs in the evening of day one and is compensated by increased production overnight, reflecting cost-optimal trade-offs between comfort and operational costs. M3 shows preheating at the start of day one, exploiting unpenalized TI in the building floor to better maintain indoor temperature in the evening without extra production. M4, on the other hand, uses TI similarly but distributes HP operation more evenly over time than M3. This smoother operation lowers supply temperatures and improves HP efficiency, while both scenarios maintain a relatively constant indoor temperature.

\section{Conclusion}
\label{sc:conclusion}
This work advances the modeling of thermal building inertia by integrating an enhanced thermal storage representation that explicitly captures floor temperature dynamics. Combined with a novel temperature-aware HP efficiency formulation, the proposed approach enables the representation of smart HP operation strategies and allows for the systematic assessment of flexibility potentials in large-scale ESOMs. The approach is demonstrated in a case study of the European energy system, identifying potential cost reductions for heating of up to~22\%. Overall, the more advanced formulations enable a more accurate representation of flexibility potentials, albeit at increased computational burden. Nevertheless, the resulting models remain computationally tractable, confirming their suitability for large-scale ESOM applications.

It should be noted that the identified flexibility potentials are derived from the perspective of a central planner. Their practical realization depends not only on the technical implementation of smart control units but also on appropriate market designs that incentivize system-optimal HP operation. A potential psychological hurdle is that individual occupants may need to relinquish control over space heating.  

Future work should extend the case study by differentiating between HP technologies, building archetypes, and heat distribution systems to obtain more robust quantitative estimates of flexibility potentials. On the modeling side, hot water buffer storage could be implemented to better capture the interplay between physical and virtual storage. 

\section*{Acknowledgment}
The authors gratefully acknowledge B. Stöckl for providing the data for the European  case study and F. C. A. Auer for his invaluable support in coding and implementing the model.

\printbibliography

\appendix 
\label{app:model_pararmetrization}
The thermodynamic heating system model is given in~\eqref{eq:thermodynamic_model}, with indices $n$ and $h$ omitted for clarity.
\begin{subequations}
\label{eq:thermodynamic_model}
\begin{align}
\mathrm{cop} &= \eta_{\mathrm{2nd}} \cdot 
\dfrac{\frac{t^R - t^S}{\ln(t^R/t^S)}}
{\frac{t^R - t^S}{\ln(t^R/t^S)} - \frac{t^{\mathrm{sc,in}} - t^{\mathrm{sc,out}}}{\ln(t^{\mathrm{sc,in}}/t^{\mathrm{sc,out}})}}
\label{eq:thermodynamic_cop}, \\[0.5em]
q^{\mathrm{hp}} &= \mathrm{cop} \cdot p^{\mathrm{hp}}
\label{eq:thermodynamic_cop_relation}, \\[0.5em]
q^{\mathrm{hp}} &= \dot{m} \cdot c_p \cdot (t^S - t^R)
\label{eq:thermodynamic_massflow}, \\[0.5em]
t^R &= t^{\mathrm{fl}} + (t^S - t^{\mathrm{fl}}) 
\cdot \exp\!\left(-U^\mathrm{tf}/(\dot{m} \cdot c_p) \right) 
\label{eq:thermodynamic_floor}.
\end{align}
\end{subequations}
Equation~\eqref{eq:thermodynamic_cop} defines the Lorenz-cycle COP as a function of sink-side supply and return temperatures $t^S$, $t^R$, source-side temperatures $t^{\mathrm{sc,in}}$, $t^{\mathrm{sc,out}}$, and second-law efficiency $\eta_{\mathrm{2nd}}$~\cite{jesper_large-scale_2021}. Heat output relates to electrical input via~\eqref{eq:thermodynamic_cop_relation}. The mass flow rate $\dot{m}$ follows from~\eqref{eq:thermodynamic_massflow} and determines the return temperature in~\eqref{eq:thermodynamic_floor} as a function of floor temperature $t^{\mathrm{fl}}$ and heat transfer coefficient $U^\mathrm{tf}$~\cite{grote_dubbel_2014}. 
The system is solved for normalized inputs $p^{\mathrm{hp}} \in [0,1]$ and floor temperatures $t^{\mathrm{fl}}$ by minimizing the supply temperature $\min \; t^S$, corresponding to maximum HP efficiency, using \texttt{SymPy} (v1.12)~\cite{sympy} and \texttt{SciPy} (v1.11.4)~\cite{scipy-NMeth}. Mass flow rates $\dot{m}$ and supply temperatures $T^{\mathrm{sup}}$ are checked for physical plausibility.

Average engineering design parameters are adopted due to limited component-specific data, while retaining full parametric flexibility. We assume $\eta^{\mathrm{2nd}} = 0.5$. For air-source HPs, standard test conditions $T^{\mathrm{sc,in}} = \SI{7}{\celsius}$ and $T^{\mathrm{sc,out}} = \SI{5}{\celsius}$ are applied. The specific heat capacity of water is set to $c_p = \SI{1.16}{\kilo\watt\hour\per\cubic\meter\per\kelvin}$ and the average COP to $\mathrm{COP}^{\mathrm{avg}} = 3.2$.
Under normalization to \SI{1}{\kilo\watt} electrical input, this yields $Q^{\mathrm{th}} = \SI{3.2}{\kilo\watt}$. At the design point, a terminal temperature difference of $\Delta T^{\mathrm{tf}} = \SI{8}{\kelvin}$ results in an effective heat transfer coefficient $U^{\mathrm{tf}} = \SI{0.4}{\kilo\watt\per\kelvin}$.


\end{document}